\documentstyle{mn}

\newif\ifAMStwofonts

\newcommand{\ra}[4]{$#1^{\mathrm{h}}#2^{\mathrm{m}}#3^{\mathrm{s}}\!\!.#4$}
\newcommand{\dec}[3]{$#1^{\circ}#2^{\prime}#3^{\prime\prime}$}

\ifoldfss
  \ifCUPmtlplainloaded \else
    \NewTextAlphabet{textbfit} {cmbxti10} {}
    \NewTextAlphabet{textbfss} {cmssbx10} {}
    \NewMathAlphabet{mathbfit} {cmbxti10} {} 
    \NewMathAlphabet{mathbfss} {cmssbx10} {} 
  \fi
  \ifAMStwofonts
    \ifCUPmtlplainloaded \else
      \NewSymbolFont{upmath} {eurm10}
      \NewSymbolFont{AMSa} {msam10}
      \NewMathSymbol{\upi}     {0}{upmath}{19}
      \NewMathSymbol{\umu}     {0}{upmath}{16}
      \NewMathSymbol{\upartial}{0}{upmath}{40}
      \NewMathSymbol{\leqslant}{3}{AMSa}{36}
      \NewMathSymbol{\geqslant}{3}{AMSa}{3E}

    \fi
  \fi
\fi 

\ifnfssone
  \newmathalphabet{\mathit}
  \addtoversion{normal}{\mathit}{cmr}{m}{it}
  \addtoversion{bold}{\mathit}{cmr}{bx}{it}
  \newmathalphabet{\mathbfit} 
  \addtoversion{normal}{\mathbfit}{cmr}{bx}{it}
  \addtoversion{bold}{\mathbfit}{cmr}{bx}{it}
  \newmathalphabet{\mathbfss} 
  \addtoversion{normal}{\mathbfss}{cmss}{bx}{n}
  \addtoversion{bold}{\mathbfss}{cmss}{bx}{n}
  \ifAMStwofonts
    \ifCUPmtlplainloaded \else
      \UseAMStwoboldmath
      \makeatletter
      \new@mathgroup\upmath@group
      \define@mathgroup\mv@normal\upmath@group{eur}{m}{n}
      \define@mathgroup\mv@bold\upmath@group{eur}{b}{n}
      \edef\UPM{\hexnumber\upmath@group}
      \new@mathgroup\amsa@group
      \define@mathgroup\mv@normal\amsa@group{msa}{m}{n}
      \define@mathgroup\mv@bold\amsa@group{msa}{m}{n}
      \edef\AMSa{\hexnumber\amsa@group}
      \makeatother
      \mathchardef\upi="0\UPM19
      \mathchardef\umu="0\UPM16
      \mathchardef\upartial="0\UPM40
      \mathchardef\leqslant="3\AMSa36
      \mathchardef\geqslant="3\AMSa3E
    \fi
  \fi
\fi 

\ifnfsstwo
  \DeclareMathAlphabet{\mathbfit}{OT1}{cmr}{bx}{it}
  \SetMathAlphabet\mathbfit{bold}{OT1}{cmr}{bx}{it}
  \DeclareMathAlphabet{\mathbfss}{OT1}{cmss}{bx}{n}
  \SetMathAlphabet\mathbfss{bold}{OT1}{cmss}{bx}{n}
  \ifAMStwofonts
    \ifCUPmtlplainloaded \else
      \DeclareSymbolFont{UPM}{U}{eur}{m}{n}
      \SetSymbolFont{UPM}{bold}{U}{eur}{b}{n}
      \DeclareSymbolFont{AMSa}{U}{msa}{m}{n}
      \DeclareMathSymbol{\upi}{0}{UPM}{"19}
      \DeclareMathSymbol{\umu}{0}{UPM}{"16}
      \DeclareMathSymbol{\upartial}{0}{UPM}{"40}
      \DeclareMathSymbol{\leqslant}{3}{AMSa}{"36}
      \DeclareMathSymbol{\geqslant}{3}{AMSa}{"3E}
    \fi
  \fi
\fi 

\ifCUPmtlplainloaded \else
  \ifAMStwofonts \else 
    \def\upi{\pi}
    \def\umu{\mu}
    \def\upartial{\partial}
  \fi
\fi

\title[The galaxies around NGC 326]{Redshift and velocity dispersion of the cluster of galaxies around NGC~326}
\author[P.N. Werner, D.M. Worrall and M. Birkinshaw]
       {P.N. Werner, D.M. Worrall and M. Birkinshaw \\
Department of Physics, University of Bristol, Tyndall Avenue, Bristol, BS8 1TL}
\date{Accepted 1999 March 22.
      Received 1999 February 22}

\pagerange{1--3}
\pubyear{1999}

\begin{document}

\maketitle


\begin{abstract}

Redshifts of several galaxies thought to be associated with NGC~326 are determined. The results confirm the presence of a cluster and find a mean redshift of $\overline{z}= 0.0477 \pm 0.0007$ and a line-of-sight velocity dispersion $ \sigma_{z} = 599\,(+230, -110)$ km~s$^{-1}$. The velocity dispersion and previously measured X-ray gas temperature of $kT \simeq 1.9$ keV are consistent with the cluster $ \sigma_{z}/kT$ relation, and NGC~326 is seen to be a slowly-moving member of the cluster.

\end{abstract}


\section{Introduction}

The radio galaxy NGC~326 ($\alpha$ = \ra{0}{58}{22}{6}, $\delta$ = \dec{26}{51}{59}, J2000.0) has been studied in soft X-rays (using the ROSAT PSPC) by Worrall, Birkinshaw \& Cameron (1995). A peak in X-ray emission coincides with the core of the radio source. A comparison of the X-ray image with the digitized Palomar Observatory Sky Survey (POSS) revealed diffuse X-ray emission asymmetrically distributed around NGC~326 and roughly coincident with the brightest objects in the North-Western quadrant of Zwicky cluster 0056.9+2636 (Zwicky \& Kowal 1968). Zw~0056.9+2636 is described as `medium compact' and `near' (i.e. with $cz < 15000$ km~s$^{-1}$, where the velocity was estimated from the apparent magnitude and diameter of the brightest cluster members), and extends over roughly a square degree with 145 galaxies brighter than 15.7 in photographic magnitude.

Worrall et al. called for optical spectral measurements to determine the redshift of the galaxies thought to be associated with the observed X-ray emitting gas, and test NGC~326's cluster membership. This paper reports spectra and redshifts for eight galaxies (including NGC~326 itself).

NGC~326 is a dumbbell system, classified by Valentijn \& Casertano (1988) as `type 1', which means that it consists of two bound D type galaxies with photographic brightnesses $<$ 1 mag apart. Throughout this paper, the two components are referred to as `core 1' and `core 2', with `core 1' being the brighter of the two (by $\sim~25$ per cent). The radio source is associated with core 1 of NGC~326.

\section{Observations}

Observations of the galaxies listed in Table~\ref{tab:obs} were carried out on the nights of 1995 February 21 and 24, using the blue channel spectrograph on the Multiple Mirror Telescope (MMT). All images were obtained on a $3072\times1024$ CCD with a 300 line mm$^{-1}$ grating.

\begin{table*}
\caption{Observational details; adapted from Table 1 in Worrall et al. (1995)} 
\label{tab:obs}
\begin{tabular}{clcccc}   \hline
Object&Name(s)&\multicolumn{2}{c}{Coordinates (J2000)}&Exp.(s)&Date observed\\ 
      &       &       R.A.           &        Dec.    &       &             \\ \hline
G1 &NGC~326, Zw 0055.7+2636 &\ra{0}{58}{22}{6} & \dec{26}{51}{59}&300& 21 Feb 1995 \\
G2 &UGC 613, Zw 0056.7+2647 &\ra{0}{59}{24}{5} & \dec{27}{03}{33}&300&$\prime\prime$ \\
G3 &MCG 04-03-030        & \ra{0}{59}{03}{6} & \dec{27}{02}{33} &300&$\prime\prime$ \\
G4 &--                   & \ra{0}{58}{47}{5} & \dec{26}{58}{40} &300& 24 Feb 1995 \\
G5 &MCG 04-03-024        & \ra{0}{58}{04}{4} & \dec{26}{53}{47} &300&$\prime\prime$ \\
G6 &--                   & \ra{0}{58}{28}{5} & \dec{26}{53}{43} &600&$\prime\prime$ \\
G7 &--                   & \ra{0}{59}{00}{4} & \dec{27}{08}{47} &480&$\prime\prime$ \\ 
G8 &--                   & \ra{0}{58}{09}{6} & \dec{26}{48}{00} &600&$\prime\prime$ \\ 
\hline
\end{tabular}
\end{table*}

The CCD images were processed using the IRAF software collection (in particular, the CCDPROC task and the APEXTRACT package), then wavelength calibrated to an accuracy of 0.8 \AA\ in $\lambda = 4000 - 9000$ \AA, using HeNeAr lamp images for each object. Those flat-field images (using a quartz lamp) taken on February 21 presented an unusual systematic pattern, which distorted the object spectra at the low and high wavelength ends; affected regions were not used in our analysis.

\section{Redshifts from emission lines}

Three galaxies showed emission lines which could be used to measure their redshift (Table~\ref{tab:zem} -- all redshifts and velocities in this paper are heliocentric). In particular, the spectrum of G2 had eight strong emission lines, whereas four weaker emission features were detected in G5. For G3, one line was found at an observed wavelength of $4575.8 \pm 3.1 $ \AA, yielding a redshift of $0.0487 \pm 0.0008$ if identified as [O III] 4363 \AA. 

Emission lines can be a valuable diagnostic tool for determining the radiation mechanism in galaxies, and hence their nature. Only for G2 are the emission lines sufficiently strong and numerous to provide useful information. The line strengths, ratios, and spatial distribution indicate that the radiation arises from gas which is photoionized by hot stars. Spectra for the galactic nucleus and for the outer edges of the galaxy show that the emission lines are slightly stronger in the off-axis spectra, indicating a likely association of the photoionising stars with spiral arms.

\begin{table}
\centering
\caption{Redshifts obtained from emission lines.} \label{tab:zem}
\begin{tabular}{lll} \hline
Object  &  Emission lines used & \multicolumn{1}{c}{$z^{\rm(em)}$} \\ \hline
G2 & H$\beta$, [O III]\ 4958, [O III]\ 5007,   & $0.0462\pm0.0001$\\
   & [O I]\ 6300, [N II]\ 6583, H$\alpha$,  & \\
   & [S II]\ 6716, [S II]\ 6731     & \\
G3 & [O III]\ 4363                             & $0.0487\pm0.0008$\\
G5 & [O III]\ 5007, H$\alpha$, [S II]\ 6716,   & $0.0457\pm0.0004$\\
   & [S II]\ 6731                                & \\ \hline
\end{tabular}
\end{table}

\section{Redshifts from absorption lines}

\begin{table}
\caption{Results of the cross-correlations.}\label{tab:zabs}
\begin{tabular}{lccrr} \hline
Object & $V^{\rm(abs)}=cz^{\rm(abs)}$ & $z^{\rm(abs)}$ & \multicolumn{1}{c}{$V_{\rm rel}$}&\multicolumn{1}{c}{$V_{\rm rel}/\sigma_{z}$}  \\
 & \multicolumn{1}{c}{km~s$^{-1}$} & & \multicolumn{1}{c}{km~s$^{-1}$} & \\ \hline
G1-1  & $14206\pm37$    &  $0.0474\pm0.0001 $ &  $ -96 $ &-0.16  \\
G1-2  & $14832\pm48$    &  $0.0495\pm0.0002 $ &  $ 500 $ & 0.83  \\  
G2    & $13840\pm41$    &  $0.0462\pm0.0001 $ &  $-447 $ &-0.75  \\
G3    & $14582\pm29$    &  $0.0486\pm0.0001 $ &  $ 262 $ & 0.44  \\
G4    & $13504\pm33$    &  $0.0450\pm0.0001 $ &  $-768 $ &-1.28  \\    
G5    & $13842\pm62$    &  $0.0462\pm0.0002 $ &  $-445 $ &-0.74  \\   
G6    & $15575\pm30$    &  $0.0520\pm0.0001 $ &  $1209 $ & 2.02  \\
G7    & $14404\pm33$    &  $0.0480\pm0.0001 $ &  $  93 $ & 0.15  \\
G8    & $13986\pm28$    &  $0.0467\pm0.0001 $ &  $-307 $ &-0.51  \\
\hline
\end{tabular}
\end{table}

We measure the absorption redshifts of the galaxies by the standard method of cross-correlating their stellar spectra with the stellar absorption-line spectrum of an object of known redshift (Tonry \& Davis 1979). The cross-correlation was carried out using the IRAF task FXCOR in the RV package. Information about the observing conditions from the observing log, and trial-and-error cross-correlations, indicated that the spectrum of G7 was a good template against which the relative velocities of the other cluster members could be measured. To provide absolute velocities, a previously processed, high-quality, normalised, zero-velocity spectrum of NGC 4486B was correlated with the spectrum of G7.

Before carrying out the cross-correlation, emission features, any remaining cosmic ray hits, and the broad telluric absorption feature around 7600 \AA\ resulting from atmospheric oxygen were deleted from the spectra.

The cross-correlation of G7 with NGC 4486B yielded a velocity for G7 of $V=14404\pm33$ km~s$^{-1}$ (corresponding redshift: $z=0.0480\pm0.0001$), with an unambiguous and narrow correlation peak, implying a satisfactory fit.

Having obtained an absolute velocity for G7, all object spectra were smoothed by the boxcar method using a 5 pixel ($\sim8$ \AA) smoothing window, and cross-correlated with the unsmoothed spectrum of G7. The spectral range 5000 -- 6500~\AA\ (containing the strong Mg feature) was best suited for the cross-correlation. In some cases, a slightly smaller range was used to avoid regions of excessive noise. The results from this procedure are shown in Table~\ref{tab:zabs}. The emission and absorption redshifts for G2, G3 and G5 are in good agreement.

\section{Velocity dispersion}

We used the data in Table~\ref{tab:zabs} to compute the mean redshift and the velocity dispersion of the cluster of galaxies using the method described in Danese, De Zotti \& di Tullio (1980). The two cores of NGC~326 were treated independently, bringing the total number of galaxies to nine. As the sample is small, uncertainties are dominated by the sampling ($\chi^{2}$) term in equation (10) of Danese et al. The measurement errors of the individual redshifts have an almost insignificant effect on the result for the velocity dispersion.

The average velocity of the cluster was found to be $c\overline{z} = 14307\pm 224$ km~s$^{-1}$, equivalent to an average redshift of $\overline{z} = 0.0477 \pm 0.0007$. The one-dimensional line-of-sight velocity dispersion was computed to be $\sigma_{z}= 599\, (+230, -107)$ km~s$^{-1}$, yielding a three-dimensional physical velocity dispersion of $\sigma= 1037\, (+427, -241)$ km~s$^{-1}$.

The last two columns of Table~\ref{tab:zabs} list the velocities of the galaxies relative to the cluster velocity, $V_{\rm rel}$, in units of km~s$^{-1}$ and in units of line-of-sight velocity dispersion $\sigma_{z}$. They show that the velocity of core 1 of NGC~326 (the radio core) is near to the cluster velocity, indicating that the radio source is not moving quickly in the gas. The velocity of G6 is found to differ from the cluster velocity by slightly more than 2$\sigma_{z}$. Leaving it out of the calculations leads to significantly reduced one- and three-dimensional velocity dispersions of $419\, (+178,-79)$ and $725\, (+329,-178)$ km~s$^{-1}$, respectively. However the justification for dropping it from the cluster membership is weak because there is a 39 per cent probability that one out of nine galaxies would lie more than $2.02\sigma_{z}$ from the mean.

\section{Discussion}

\subsection{Earlier measurements of the velocity of NGC~326}

Three previous redshift measurements for NGC~326 are compared with our results in Table~\ref{tab:vcomp}.

The earliest measurement, from H and K absorption features in the two components of NGC~326, appears in a list of 51 galaxies identified with radio sources in the 4C catalogue where NGC~326 is 4C26.03 (Sargent 1973). The absence of errors on these velocities makes it difficult to compare them with our results.
  
De Vaucouleurs et al. (1991) gives a  velocity of $14138\pm59$ km~s$^{-1}$, equivalent to a redshift of $z = 0.0472\pm 0.0002$, without reference to the dumbbell nature of NGC~326. This agrees with our result for core 1, the brighter of the two components.
  
A more recent redshift measurement was made by Davoust \& Consid\`{e}re (1995) in a study of the kinematics of galaxy pairs. Their lower and upper velocities for the two components of NGC~326, $14313 \pm 36$ and $14822 \pm 25$ km~s$^{-1}$, are in adequate agreement with our velocities and suggest no significant systematic error in our velocity scale.

\begin{table}
\caption{Comparison of measured velocities} \label{tab:vcomp}
\begin{tabular}{lclll} \hline
Authors & Year & \multicolumn{1}{c}{$cz_{\rm core\,1}$} & \multicolumn{1}{c}{$cz_{\rm core\,2}$}  & \multicolumn{1}{c}{$cz_{\rm 2}-cz_{\rm 1}$} \\ 
 & & \multicolumn{1}{c}{km~s$^{-1}$} & \multicolumn{1}{c}{km~s$^{-1}$} & \multicolumn{1}{c}{km~s$^{-1}$}\\
\hline
Sargent         & 1973& $14120^{\,a}$  &  $14840^{\,a}$  &  $720$ \\
de V. et al.    & 1991& $14138 \pm59$  &---              &--- \\
D. \& C.        & 1995& $14313 \pm36$  &  $14822\pm 25$  & $ 509\pm 44 $ \\
{This paper}    &     & $14206 \pm37$&$14832 \pm 48$   &$549\pm38^{\,b}$ \\ \hline
\end{tabular}
$^{a}$\ listed as redshift (0.0471 and 0.0495, respectively)

$^{b}$\ obtained by cross-correlating the two cores rather than by calculating the difference $cz_{\rm 2}-cz_{\rm 1}$
\end{table}

\subsection{Velocity dispersion and X-ray temperature}

Previous authors have found a relation between the velocity dispersion of a cluster of galaxies and the X-ray temperature of the gas within the cluster. Slightly different but consistent expressions for this relation have been derived by Lubin \& Bahcall\, (1993), Bird, Mushotzky \& Metzler (1995) and Girardi et al.\, (1996). Girardi et al.'s expression,\ $(\sigma_{z}/$km~s$^{-1})=10^{(2.53\pm0.04)}\times(T/$keV$)^{(0.61\pm0.05)}$, leads to a predicted velocity dispersion $\sigma_{z} = 501\ (+232, -113)$ km~s$^{-1}$ based on the temperature of $kT = 1.9\, (+0.9,-0.4)$ keV ($1\sigma$ error) found by Worrall et al. (1995). The result is in good agreement with our measured one-dimensional velocity dispersion, supporting the argument that the X-ray gas detected by Worrall et al. (1995) is associated with the cluster containing NGC~326.

\subsection{Velocity dispersion and cluster richness}

In their study of 43 clusters, Danese et al. (1980) found a correlation between Abell's (1958) richness class and the physical velocity dispersion, as might be expected if the galaxy population is a measure of total mass. For richness classes R~=~0, 1 and 2, the mean three-dimensional velocity dispersions were found to be $820 \pm 64$ km~s$^{-1}$, $1400 \pm 210$ km~s$^{-1}$ and $1760\pm 120$ km~s$^{-1}$, respectively. From our (three-dimensional) velocity dispersion of $\sigma = 1037\ (+427, -241)$ km~s$^{-1}$, it can be deduced that the cluster studied in this paper falls between the richness classes 0 and 1. A visual inspection of the optical sky survey image suggests that this appears to be a reasonable richness estimate for the Western side of the cluster Zw 0056.9+2636.

\section{Conclusions}

In his 1968 catalogue, Zwicky indicated a population of 145 for the cluster Zw~0056.9+2636. Our velocity dispersion (an indication of Abell richness class) suggests that the galaxies detected by Zwicky are not all physically associated, because our measured velocity dispersion is too low. This conclusion is consolidated by the findings of Worrall et al. (1995), who detected X-ray gas only in the Western part of the cluster. Subdividing the cluster into two halves, the following can be concluded:

\begin{itemize}

\item{In the \emph{North-Western} part of Zwicky cluster Zw~0056.9+2636, dominated by the dumbbell radio galaxy NGC~326, the physical and line-of-sight velocity dispersions are consistent with the cluster richness and the X-ray gas temperature. The gas  morphology, which is asymmetric but peaked at NGC~326, and the modest temperature suggest that this cluster segment is rather young. The radio core of NGC~326 is slowly moving relative to the mean velocity of the galaxies in the cluster.}

\item{No redshift or velocity dispersion information is available for the \emph{South-Eastern} end of Zw~0056.9+2636, which appears to contain much of the galaxy count, but little hot gas. The physical relationship of this galaxy concentration with the cluster around NGC~326 remains unclear.}

\end{itemize}

\section*{Acknowledgments}

We thank Glenn Baggley, Claire Halliday and Brian McNamara for assistance with the optical observing.



\end{document}